\documentclass[journal=jacsat,manuscript=article]{achemso}

\usepackage{chemformula} 
\usepackage[T1]{fontenc} 
\usepackage{isotope}
\usepackage[caption=false]{subfig}

\title{Static and dynamic disorder effects on the magnetism in FeRh}	
		\author{Benedikt Eggert}
		\email{Benedikt.Eggert@uni-due.de}
		    \affiliation{Faculty of Physics and Center for Nanointegration Duisburg-Essen (CENIDE), University of Duisburg-Essen, Lotharstr. 1, D-47057, Duisburg, Germany}
		\author{Alexander Schmeink}
		    \affiliation{Institute for Ion Beam Physics, Helmholtz-Zentrum Dresden-Rossendorf, Bautzner Landstr. 400, 01328 Dresden, Germany}
		\alsoaffiliation{Dresden University of Technology, Helmholtzstr. 10, 01069 Dresden, Germany}
        \author{Johanna Lill}
            \affiliation{Faculty of Physics and Center for Nanointegration Duisburg-Essen (CENIDE), University of Duisburg-Essen, Lotharstr. 1, D-47057, Duisburg, Germany}
         \author{Maciej Oskar Liedke}
            \affiliation{Institute for Radiation Physics, Helmholtz-Zentrum Dresden-Rossendorf, Bautzner Landstr. 400, 01328 Dresden, Germany}
        \author{Ulrich Kentsch}
		    \affiliation{Institute for Ion Beam Physics, Helmholtz-Zentrum Dresden-Rossendorf, Bautzner Landstr. 400, 01328 Dresden, Germany}
        \author{Maik Butterling}
            \affiliation{Institute for Radiation Physics, Helmholtz-Zentrum Dresden-Rossendorf, Bautzner Landstr. 400, 01328 Dresden, Germany}
        \author{Andreas Wagner}
            \affiliation{Institute for Radiation Physics, Helmholtz-Zentrum Dresden-Rossendorf, Bautzner Landstr. 400, 01328 Dresden, Germany}
        \author{Sakura Pascarelli}
    		\affiliation{European Synchrotron Radiation Facility (ESRF), BP 220, 71 Avenue des Martyrs, 38000 Grenoble France}
		\author{Kay Potzger}
		    \affiliation{Institute for Ion Beam Physics, Helmholtz-Zentrum Dresden-Rossendorf, Bautzner Landstr. 400, 01328 Dresden, Germany}
		\author{J\"urgen Lindner}
		    \affiliation{Institute for Ion Beam Physics, Helmholtz-Zentrum Dresden-Rossendorf, Bautzner Landstr. 400, 01328 Dresden, Germany}
		\author{Thomas Thomson}
            \affiliation{School of Computer Science, The University of Manchester, Oxford Road, Manchester M13 9PL, United Kingdom}
		\author{J\"urgen Fassbender}
		    \affiliation{Institute for Ion Beam Physics, Helmholtz-Zentrum Dresden-Rossendorf, Bautzner Landstr. 400, 01328 Dresden, Germany}
	    	\alsoaffiliation{Dresden University of Technology, Helmholtzstr. 10, 01069 Dresden, Germany}
        \author{Katharina Ollefs}
		    \affiliation{Faculty of Physics and Center for Nanointegration Duisburg-Essen (CENIDE), University of Duisburg-Essen, Lotharstr. 1, D-47057, Duisburg, Germany}
		\author{Werner Keune}
		    \affiliation{Faculty of Physics and Center for Nanointegration Duisburg-Essen (CENIDE), University of Duisburg-Essen, Lotharstr. 1, D-47057, Duisburg, Germany}
		\author{Rantej Bali}
		    \affiliation{Institute for Ion Beam Physics, Helmholtz-Zentrum Dresden-Rossendorf, Bautzner Landstr. 400, 01328 Dresden, Germany}
		\author{Heiko Wende}
		    \affiliation{Faculty of Physics and Center for Nanointegration Duisburg-Essen (CENIDE), University of Duisburg-Essen, Lotharstr. 1, D-47057, Duisburg, Germany}

\title{Magnetic response of FeRh to static and dynamic disorder}	

\begin{document}

\begin{tocentry}

    \includegraphics[width=.71\linewidth]{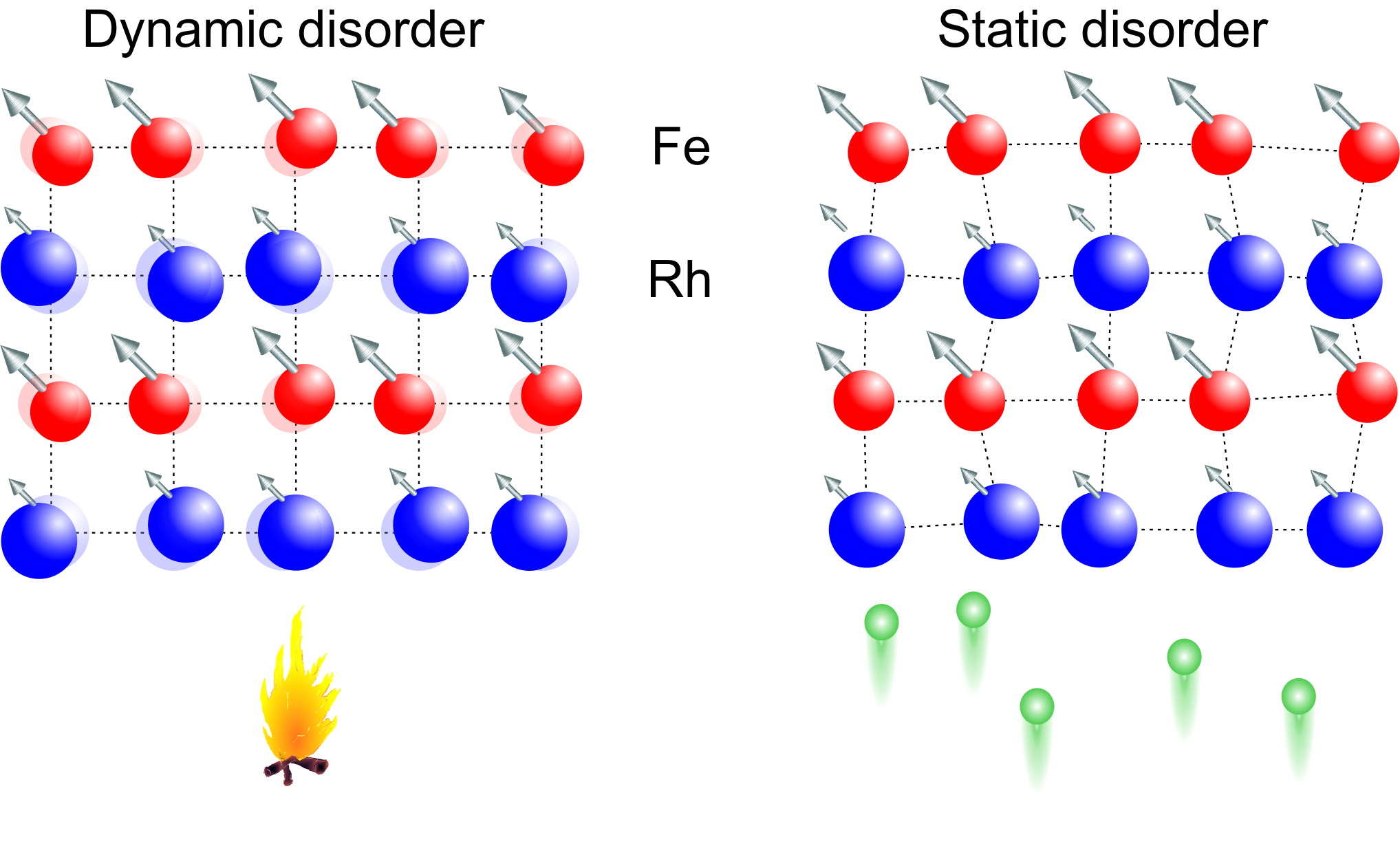}

\end{tocentry}

\begin{abstract}
\noindent
Changes of the magnetic and crystal structure on the microscopic scale in 40\,nm FeRh thin films have been applied to investigate the phenomena of a disorder induced ferromagnetism at room temperature initiated through light ion-irradiation with fluences up to 0.125\,Ne$^+$/nm$^{-2}$. Magnetometry shows an increase of magnetic ordering at low temperatures and a decrease of the transition temperature combined with a broadening of the hysteresis with rising ion fluence. \isotope[57]Fe M\"ossbauer spectroscopy reveals the occurrence of an additional magnetic contributions with an hyperfine splitting of 27.2\,T - identical to that of ferromagnetic B2-FeRh. The appearance of an anti-site Fe-contribution can be assumed to be lower than 0.6\,Fe-at\%, indicating that no change of the chemical composition is evident. The investigation of the local structure shows an increase of the static mean square relative displacement determined by X-ray absorption fine structure spectroscopy, while an increase of the defect-concentration has been determined by positron annihilation spectroscopy. From the changes of the microscopic magnetic structure a similarity between the temperature induced and the structural disorder induced ferromagnetic phase can be observed. These findings emphasize the relationship between magnetic ordering and the microscopic defect structure in FeRh.
\end{abstract}

\section{Introduction}

Equiatomic FeRh is well-known for the unique properties, including a first-order metamagnetic isostructural transition from a low tempe\-rature antiferromagnetic (AFM) to a high temperature ferromagnetic (FM) phase at the transition temperature $T_{tr}$ of approximately 370\,K. Potential use cases to deploy this metamagnetic transition include heat-assisted magnetic recording (HAMR) \cite{Thiele2003}, magnetotransport\cite{Sharma2011}, antiferromagnetic spintronics \cite{Marti2014} and magnetic refrigeration \cite{Cooke2012,Zhou2013,Gutfleisch2016}.  For this purpose, efforts are made to engineer the hysteresis, for example by growing epitaxial thin films on different substrates \cite{Witte2016,Witte2016a,Ceballos2017}, varying the strain affected volume by modifying the film-thickness \cite{Ostler2017,Barton2017} or doping the system to shift the phase transition towards lower or higher temperatures \cite{Barua2014,Jiang2016}. During this metamagnetic phase transition of equiatomic FeRh the local Fe and Rh moments increase from $\pm$ 3$\,\mu_B$ and 0$\,\mu_B$ in the AFM phase to 3.3$\,\mu_B$ and 1.0$\,\mu_B$, respectively, in the FM phase, accompanied by a volume increase of 1\%\cite{Shirane1963,Shirane1963a,Shirane1964}. The occurrence of the local Rh moment in the FM phase is important due to the stabilizing effect shown in Monte-Carlo simulations by Gruner et al. \cite{Gruner2003}. From an experimental point of view the instability of the Rh moment could be observed by neutron diffraction \cite{Shirane1963a,Shirane1964} or in XMCD measurements at the Rh L$_{2,3}$ \cite{Chaboy1999} and M$_{2,3}$ edges \cite{Stamm2008}.

Apart from the isostructural phase transition, B2-FeRh also possesses a disorder-induced phase transition, which can be driven, for example, by ion or laser irradiation \cite{Ehrler2018}. This process has been known for FeRh and was initially shown by Iwase et al. \cite{Iwase2007}, where irradiating a bulk FeRh sample with Ni, Kr, Xe or Au ions, lead to a finite orbital polarization measured by XMCD at the Fe K edge below room temperature. Systematic disordering is achieved via irradiation with light noble gas ions, such as He$^+$ or Ne$^+$ with sufficient energy to cause displacements of the Fe and Rh atoms from their ordered sites without artificially doping the system with materials possessing free electrons. Recently, similar investigations have been performed for FeRh thin films irradiated with He$^+$ \cite{Bennett2017} and Ne$^+$ \cite{Heidarian2015,Cervera2017} showing, for low irradiation fluence, an increase of the magnetization at low temperature. Previously, only microscopic investigations of the disorder-induced ferromagnetic phase in FeRh were performed with measurement techniques, which integrate all magnetic contributions making it impossible to separate contributions from the different crystallographic sites or phases in metallic systems.

In comparison, a similar effect occurs in FeAl, where the recombination of vacancies in ion irradiated B2-ordered FeAl thin films leads to the formation of a disordered bcc structure (A2 structure) with ferromagnetic ordering at room temperature\cite{Bali2014,Roeder2015}. EXAFS measurements at the Fe K edge reveal the formation of Fe and Al-rich regions with increasing irradiation fluence, accompanied by an expansion of the lattice \cite{Torre2018}. 

Additional investigations of this system showed the existence of a new monoclinic ground state\cite{Kim2016,Aschauer2016,Wolloch2016} , due to a lattice instability. It was shown, that the low temperature AFM phase is softer compared to the FM phase, as it was also demonstrated in a temperature-dependent EXAFS study performed by Wakisaka et al. \cite{Wakisaka2015}. In a recent work of Keavney et al., it could be shown by a combination of XMCD-PEEM and nano-XRD measurements, that the magnetostructural phase transition exhibits a defect-driven domain nucleation behaviour \cite{Keavney2018}. Similar effects of an inhomogeneous phase transition have been observed in a TEM study of Gatel et al. \cite{Gatel2017}, where the film-surface and the film-substrate interface have a lower transition temperature than the centre of the film. Saidl et al.\cite{Saidl2016} observed that the optical properties of different microscopic regions possess different transition temperatures leading to a distribution of transition temperatures $T_{tr}$. Furthermore, these results can be used to explain a stable FM phase located at the surface, which has been shown by Pressacco et al. \cite{Pressacco2016}. 

In this work, we investigate the effect of static disorder compared to dynamic disorder on the metamagnetic phase transition of FeRh. Dynamic disorder is defined as the motion of nuclei for example by means of zero-point vibrations at low temperatures or at higher temperatures by lattice vibrations (phonons),leading to a oscillating atomic displacement. This kind of disorder is strongly dependent on temperature. In contrast, static disorder is temperature independent and is created, for example, at the time of the crystallization process during sample preparation, effectively minimizing the formation energy. Static disorder can, for example, occur in the form of mono-vacancies, vacancy cluster, anti-site contributions (nuclei on a different crystallographic site) or a grain boundary. These kinds of defects lead to a change of the microscopic physical properties, for example, an anti-site contribution or a vacancy leads to a change of the exchange interaction. This can lead to the formation of ferromagnetism in an otherwise non ferromagnetic system\cite{Reddy2001}. On the other hand, a grain boundary or void can lead to a distortion of the lattice, effectively varying the nearest neighbour distance. In a time-averaged microscopic picture, these static variations of the nearest neighbour distance have the same effect as lattice vibrations leading to a blurred distribution of bond lengths. 

For a detailed investigation of the difference or similarities of dynamic and static disorder, it is necessary to investigate the system on a microscopic scale by considering the  electronic structure, local structure and the defect structure. We combine magnetometry measurements with conversion electron M\"ossbauer spectroscopy (CEMS) in order to relate changes in macroscopic magnetization to the local magnetic properties and the electronic structure of the system. With this approach, we can separate changes of the electronic structure on the microscopic scale and correlate these with changes of the macroscopic scale. Also, changes of the local structure by element-specific extended X-ray absorption spectroscopy at the Fe K edge will be presented, while an interlink between defects and the disordering process will be highlighted by positron annihilation spectroscopy (PAS). We show that for an ion irradiation fluence below $0.125\,\mathrm{Ne}^+/\mathrm{nm}^2$ a similarity of the AFM-FM transition occurs compared to a temperature-induced phase transition.

Furthermore, the results of this work can be applied for example to Heusler alloys, which can be used for magnetic refrigeration \cite{Weise2018} or spintronics applications\cite{Karel2017}, but posses an inherent influence of site disorder concerning the structural and magnetic order \cite{Schleicher2017}.

\section{Results}

\subsection{Macroscopic magnetic properties}

\begin{figure}[htp]
	\centering
	\includegraphics[width=\linewidth]{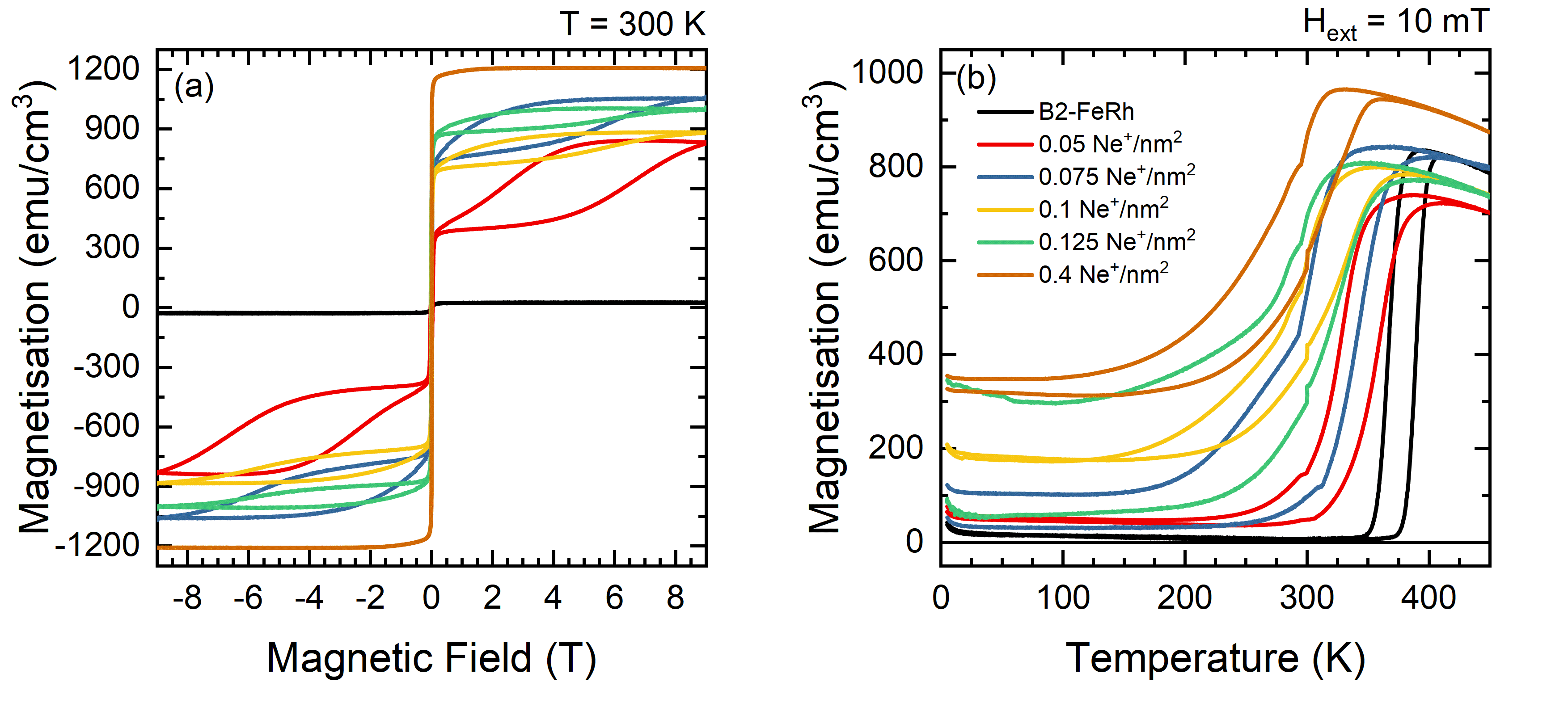}
	\caption{Macroscopic magnetization as a function of magnetic field (a) and temperature (b) for different irradiation fluences. The field dependent measurements were performed at 300\,K, while for the temperature dependent measurements an external magnetic field of 10\,mT was applied. The same color code is valid for (a) and (b)}
	\label{fig:magnetometry}
\end{figure}

Field dependent measurements performed at 300\,K (figure \ref{fig:magnetometry}a) shows, for a MBE grown 40\,nm FeRh thin film, with a residual A1 phase, a small saturation magnetization of 25$\mathrm{\,emu/cm^3}$ consistent with an AFM-ordered system with a small fraction of non-compensated Fe-spins particulary at the surface \cite{Pressacco2016}. For the irradiated samples, a hysteresis divided into two parts is observed. In small applied magnetic fields a hysteresis occurs with a coercive field below 50\,mT , while for higher magnetic fields a second loop occurs which closes in the case of applying a field of 9\,T. This loop can also be observed in the ordered samples at temperatures below the phase transition by applying a large magnetic field (e.g. $H \approx 11$\,T for $T\approx 300\mathrm{\,K}$) \cite{Zhou2013,Maat2005} and can be interpreted as the field induced isostructural AFM-FM phase transition\cite{Ibarra1994,Vries2013}. Temperature-dependent magnetization measurements show, that for the initial B2-ordered sample the first-order phase transition occurs at 370\,K with a symmetric thermal hysteresis width of 10\,K (see figure \ref{fig:magnetometry}b). The increase of the magnetization at low temperatures originates from defects and impurities \cite{Ricci2003} in the MgO substrate. With increasing ion irradiation fluence, the initial sharp transition occurs at lower temperature and the hysteresis width increases, while additionally, the magnetization at low temperatures increases.

\subsection{Microscopic magnetic properties}
\label{sec:CEMS}
\begin{figure}[htp]
    \centering
    \includegraphics[width=\linewidth]{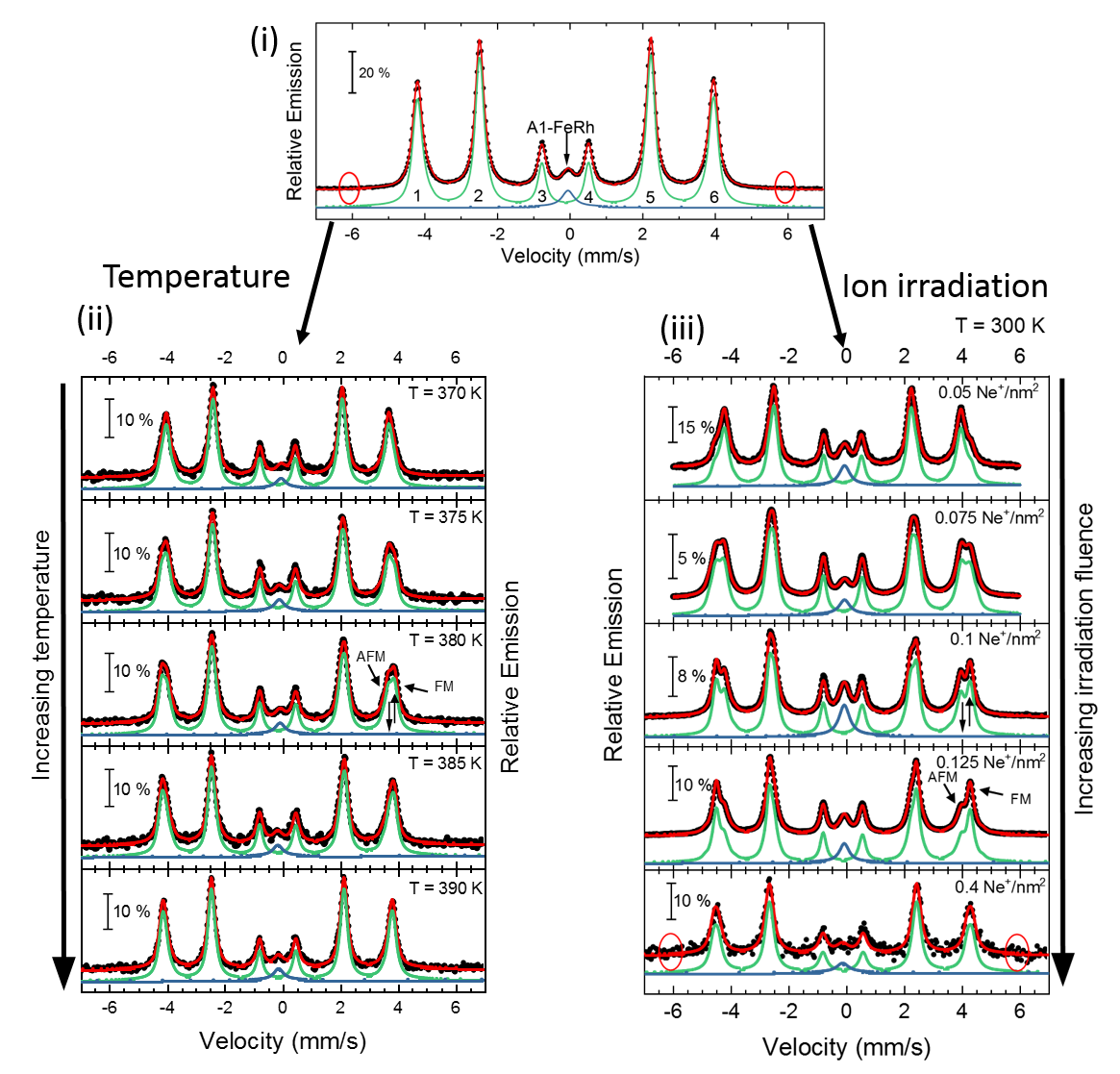}
    \caption{Zero-field Conversion Electron M\"ossbauer Spectroscopy (CEMS) results for  40\,nm thick FeRh thin film. (i) Room temperature M\"ossbauer spectrum and corresponding least-squares fit for the B2-FeRh thin film using a hyperfine-field distribution $p\left(B_{hf}\right)$ for the sextet (green) and a Lorentzian single line for the central weak singlet (blue). The corresponding nomenclature of the different sextet lines is shown. (ii) Spectra for different temperatures across the phase transition in the AFM-FM coexistence region. The corresponding measurement temperature is presented in each layer. (iii) Spectra for samples irradiated with different ion fluences varying from 0.05 up to 0.4 Ne$^+$/nm$^2$ with an ion energy of 25\,keV. In all graphs, the blue subspectra describe a paramagnetic secondary phase, while the green subspectra illustrates the contribution of hyperfine fields. The obtained hyperfine field distribution $p\left(B_{hf}\right)$ for the different measurements can been seen in fig. \ref{fig:Contour-Plots}. The expected positions of sextets line 1 and 6 caused by anti-site Fe \cite{Shirane1963,Shirane1963a} is highlighted by red circles in Fig. \ref{fig:CEMS}(i). Details of the fitting procedure are given in the text.}
    \label{fig:CEMS}
\end{figure}

M\"ossbauer spectroscopy probes slight deviations in the energy levels of the \isotope[57]Fe-nuclei due to the presence of hyperfine interactions, and measures the valence state of iron and the orientation of the Fe-spin relative to the incident $\gamma$-ray. Since the hyperfine field may be roughly proportional to the magnetic moment\cite{Vincze_Kaptas_Kemeny_Kiss_Balogh_1994} changes of the microscopic Fe moment can be probed. Room temperature zero-field measurements show two distinct states in the B2-FeRh film  (see fig \ref{fig:CEMS}i). The majority phase can be described by a magnetic split sextet state with peaks of the lines 1 and 6 at -4.2\,mm/s and +4.1\,mm/s respectively, corresponding to a hyperfine field $ B_{hf} = 25.4\,\mathrm{T}$. This is in good agreement with different M\"ossbauer investigations of AFM B2-ordered FeRh \cite{Shirane1963,Bordel2012}. The intensity ratio of lines 2 (4) and 3 (5), also referred to as $A_{2,3}$-ratio describes the average angle $\theta$ between Fe-spin and incident $\gamma$-ray, by the formula.
\begin{equation}
	A_{2,3}=\dfrac{I_2}{I_3} = \dfrac{4\sin^2(\theta)}{1+\cos^2(\theta)}.
\end{equation} 
Therefore, the $A_{2,3}$-ratio exhibits values between 0 ($\Theta=0^{\circ}$ with spin orientation out of plane) and 4 ($\Theta=90^{\circ}$ with spin orientation in-plane). From the performed measurements, we determine an $A_{2,3}$-ratio of 3.9, corresponding to an almost in-plane spin orientation. The second contribution, (a single peak at about -0.1\,mm/s) can be attributed to A1-FeRh \cite{Shirane1963,Chirkova2016,Chirkova2017}, which exhibits paramagnetic ordering at RT. Both spectral contributions show a small negative isomer shift $\delta_{iso}$ of -0.01\,mm/s, close by that of bulk bcc-Fe, which represents the metallic character of the sample and no oxide contribution ($\delta_{iso}(\mathrm{oxide})>0.3\mathrm{\,mm/s}$) is present. Upon rising temperatures (fig \ref{fig:CEMS}ii), the hyperfine splitting of the sextet decreases and close to the transition temperature at 380\,K a third contribution appears, which indicates the isostructural AFM-FM transition. As is seen for temperatures below 380\,K a decrease of the hyperfine field occurs, expected for an AFM-system approaching its N\'{e}el tem\-pe\-ra\-ture $T_N$ and at 375\,K the third phase occurs, leading to an increase of the average hyperfine field by 1.4\,T. The third phase with the increased hyperfine field is assigned to the FM phase of B2-FeRh \cite{Shirane1963}. The magnetic splitting of this third magnetic state also decreases upon heating. By u\-sing a Brillouin-function to describe the temperature dependence of the average $B_{hf}$, one can determine an extrapolated N\'{e}el temperature $T_N = 615 \pm 17\mathrm{\,K}$ for the first phase and a Curie temperature $T_C=662\pm 13 \mathrm{\,K}$ for the third phase, the latter being in reasonable agreement with the Curie temperature obtained by magnetometry on thin films ($T_C=670\mathrm{\,K}$) \cite{Heidarian2017} and bulk materials ($T_C=675\mathrm{\,K}$) \cite{Kouvel1962}.


In the chemically disordered system after irradiation with 25\,keV Ne$^+$ with a small ion fluence (figure \ref{fig:CEMS}iii), changes in the magnetically ordered state are observed. From the initially ordered state, it is seen that an additional spectrum arises leading to a broadening of lines 2 and 5 combined with an additional fine structure at lines 1 and 6 corresponding to a hyperfine field of 27.4\,T. With increasing irradiation fluence the relative spectral area of this phase increases, which leads to an average hyperfine field of 27.4\,T for the maximum irradiated sample. Hence, anti-site Fe (Fe on an initial Rh site with $B_{hf}(Fe_{II}=38\,\mathrm{T}$) is not created, as one can observe for a Fe-rich B2-FeRh sample, or also the formation of the A2 phase \cite{Shirane1963} ($B_{hf}(\mathrm{A2})=35\,\mathrm{T}$) does not occur. As discussed for the Mössbauer results, one can see that the samples have different amounts of A1-FeRh (blue central singlet), showing an inhomogeneous distribution of this impurity phase \cite{Chirkova2017}, while for the present low irradiation fluences a formation of additional A1-FeRh does not occur\cite{Heidarian2015,Cervera2017,Bennett2017}. 

\begin{figure}[htp]
\centering
\includegraphics[width=\linewidth]{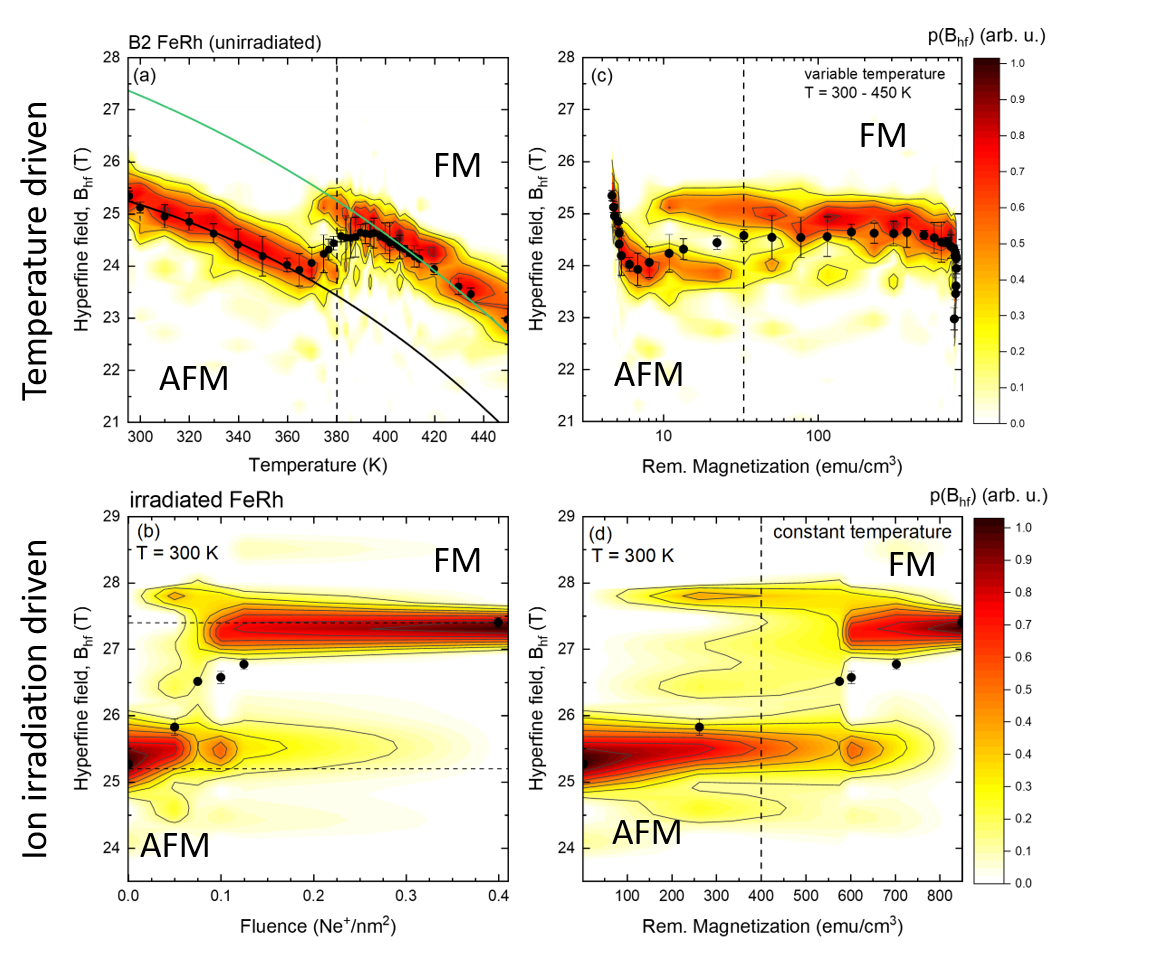}
	\caption{Hyperfine field distribution $p\left(B_{hf}\right)$ (color code) obtained from zero-field conversion electron M\"ossbauer spectroscopy for a 40\,nm FeRh thin film. Subfigure (a) shows the changes of the hyperfine field distribution across the magnetostructural phase transition and subfigure (b) illustrates the hyperfine field distribution at 300\,K for different disordered states obtained by ion irradiation with 25\,keV Ne$^+$ with different fluences. Subfigures (c) and (d) presents the $p\left(B_{hf}\right)$  distribution as a function of the macroscopic (remanent) magnetization obtained from temperature dependent (c) or field dependent measurements (d) shown in Figure \ref{fig:magnetometry}. The value of the average hyperfine field $\langle B_{hf}\rangle$ for each measurement is indicated with a black dot.}\label{fig:Contour-Plots}
\end{figure}

All of the measured magnetically split spectra were described by a hyperfine field distribution, which is presented for the temperature-driven phase transition and irradiation induced transition in fi\-gure \ref{fig:Contour-Plots}. By defining the macroscopic remanent magnetization as an order parameter of the individual phase transition, one can compare the changes of the microscopic \isotope[57]Fe hyperfine field for a system with a thermally driven phase transition (Figure \ref{fig:Contour-Plots}a) to a sys\-te\-ma\-tically structural disordered system (Figure \ref{fig:Contour-Plots}c). The plots for the dependence of the microscopic hyperfine field as a function of the macroscopic magnetisation are shown in Figure \ref{fig:Contour-Plots}b and \ref{fig:Contour-Plots}d. From this comparison we find that by a systematic increase of the structural disorder, a FM-phase with a hyperfine field of 27.4\,T is induced (Figure \ref{fig:CEMS}iii), which corresponds to the hyperfine field found at room temperature for ferromagnetic B2 ordered Fe$_{51}$Rh$_{49}$ \cite{Shirane1963,Shirane1963a,Shirane1964}. These measurements demonstrate, that the metamagnetic isostructural phase transition can be driven by ion irradiation with low fluences, as the changes of the microscopic moment as a function of the macroscopic remanent magnetisation show a similar trend compared to a temperature-driven system (figure \ref{fig:Contour-Plots}). 

\subsection{Microscopic local structure}

\begin{figure}[htp]
	\centering
	\includegraphics[width=\linewidth]{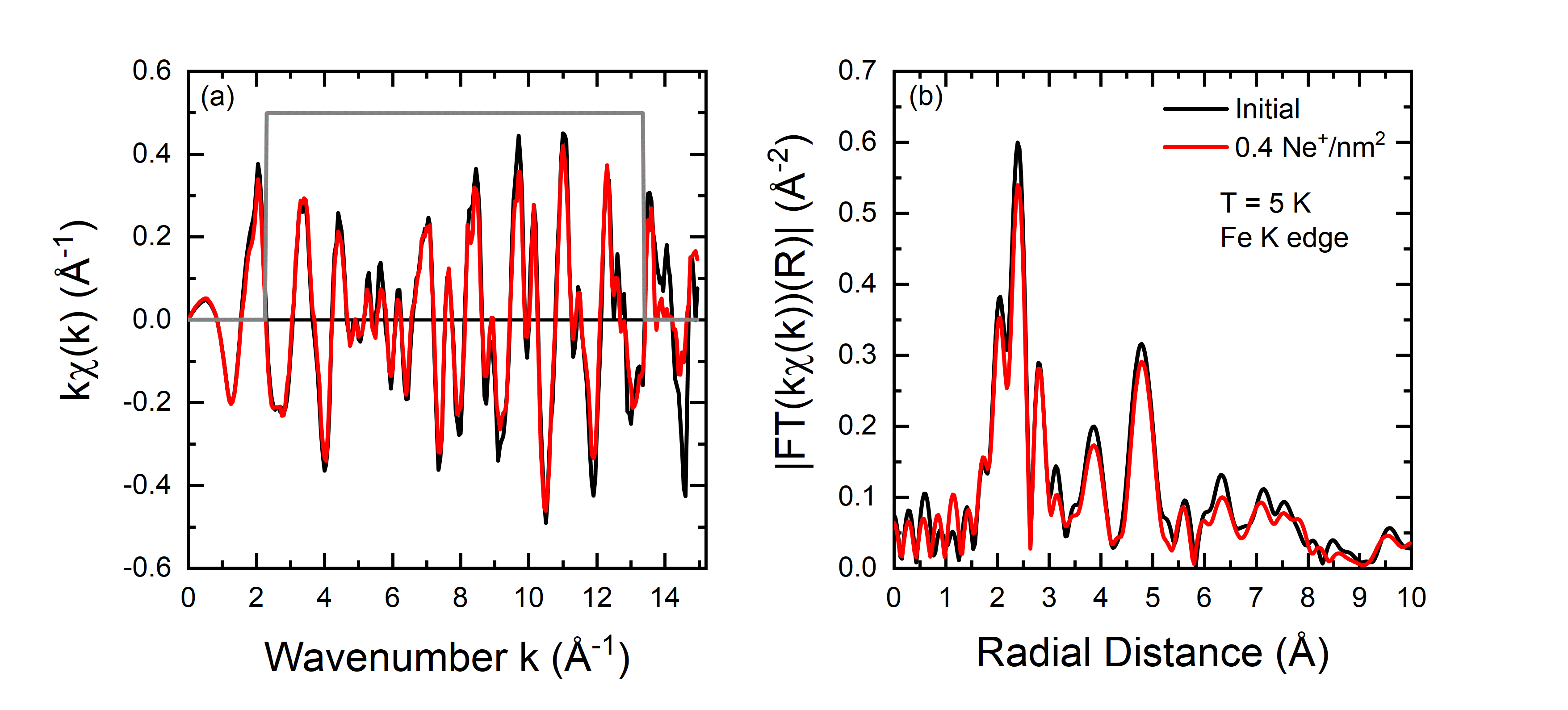}
   \caption{(a) Fe K edge k weighted EXAFS oscillations $k\chi\left(k\right)$  for an initial FeRh (black) thin film and the sample irradiated with 0.4\,Ne$^+$/nm$^2$ (red). (b) Corresponding Fourier transforms $\left|FT\left(k\chi\left(k\right)\right)\right|$. Details concerning the Fourier transformation are given in the text. Measurements have been performed at $T = 5\mathrm{\,K}$.}
    \label{fig:EXAFS}
\end{figure}
EXAFS measurements have been performed at the Fe K edge at low temperatures ($T = 5$\,K) to identify changes of the microscopic local structure. For the analysis of the measured spectra and to resolve the fine structure $\chi\left(k\right)$ the DEMETER package tool \cite{Ravel2005} has been used. For the Fourier transformation of the signal a Kaiser-Bessel window function has been used  starting at $k_{min}=2.3\,\mathring{A}^{-1}$ and ending at $k_{max}=13.3\,\mathring{A}^{-1}$ ($\varDelta k = 11\,\mathring{A}^{-1}$) with a width of $dk=0.1\,\mathring{A}^{-1}$. The Figure \ref{fig:EXAFS}a shows a small decrease of the amplitude in the fine structure, while no clear change of the oscillations can be discovered by comparing an as-grown film with a film irradiated with 0.4 Ne$^+$/nm$^2$. 

In comparison, a similar $\chi\left(k\right)$ has been observed for FeRh in the work of Wakisaka et al. \cite{Wakisaka2015} by temperature-dependent EXAFS measurements at Fe and Rh K edge. The amplitude of the Fourier Transform is illustrated in Figure \ref{fig:EXAFS}b for the two different states. The profiles are very similar, with an overall slight decrease in the intensity of the peaks. The first peak between 1.5-3$\,\text{\AA}$ originates from back-scattering from the nearest neighbour Rh shell. The dip at 2$\,$\AA \cite{McKale1988,Hanham2001} is due to the Ramsauer Townsend effect, visible on the $\chi\left(k\right)$ as a marked minimum in the amplitude near 5$\,\mathring{A}^{-1}$. 

   \subsection{Characterisation of open volume defect concentration }
  
   \begin{figure}[ht]
       \centering
       \subfloat{\includegraphics[width=.8\linewidth]{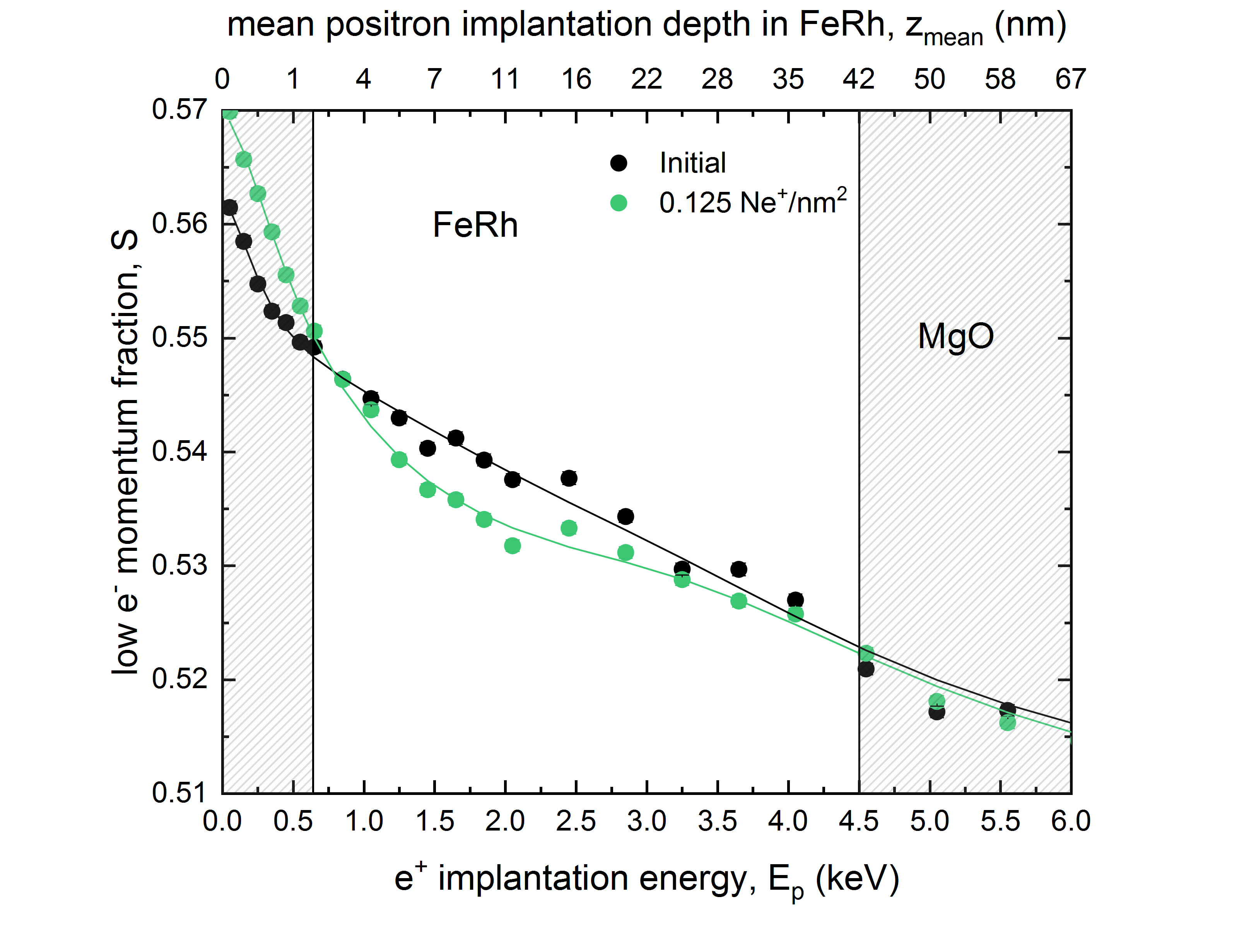}}\\
       \caption{Line shape parameter $S$ of the annihilation as a function of the incident positron energy obtained by DB-VEPAS. Theoretical fits of the $S(E)$ spectrum are presented for the as-grown and irradiated samples with a fluence of 0.125\,Ne$^{+}$/nm$^{2}$.}
              \label{fig:Positron}
   \end{figure}
Doppler broadening variable energy positron annihilation spectroscopy (DB-VEPAS) probes the open volume defects in a solid, where one detects changes of width and intensity of the 511\,keV positron annihilation line. The $S$-parameter is the fraction of positrons annihilating with low momentum valence electrons. It represents vacancy type defects and their concentration. Plotting the calculated S parameter as a function of positron implantation energy, $S(E)$, provides depth dependent information on the defect concentration \cite{Clement1996}. For the analysis of positron diffusion length, $L_+$, which is inversely proportional to defect concentration, the VEPFit code \cite{vanVeen1995} has been utilised, which permits to fit $S(E_p)$ curves for multilayered systems and to acquire thickness, $L_+$, and the specific S-parameters for each layer within a stack. $L_+$ has been calculated as 19.7(1), 7.27(4), and 5.68(4)\,nm for the as-grown, irradiated with 0.1\,Ne$^+$/nm$^{2}$, and 0.125\,Ne$^+$/nm$^{2}$ films, respectively. The thickness of the film was fixed to 42.8\,nm for all three samples, utilizing a material density of $\rho_{FeRh}=9.76$\,g/cm$^{3}$ for FeRh and $\rho_{MgO}=3.6$\,g/cm$^{3}$ for MgO respectively. $L_+$ for MgO was fixed to 35\,nm for all the samples (the fitted values: 33-37\,nm). The as-grown sample, because of a large $L_{+}$, has a relatively low defect concentration nearly as low as the MgO crystal. Ion irradiation produces damage to the film structure introducing vacancy like defects. Most likely, the size of defects remains the same and only the defect concentration increases (by a factor of 4 for the largest fluence). DB-VEPAS clearly shows an increase in the defect concentration, Figure \ref{fig:Positron}. This implies that the fraction of the open volumes, the empty spaces in the lattice increase with increasing fluence. However, the defect type, whether purely static disorder or vacancy like defects, cannot be ascribed from the Doppler Broadening results alone. Nevertheless, the increase of the open volumes is consistent with the increased static disordering observed in the ion-irradiated films, as determined from the EXAFS measurements.

\section{Discussion}

From the performed EXAFS measurements it is evident, that a change in the chemical composition of the nearest neighbour shell upon ion irradiation is not detected. A variation of the chemical composition can give rise to the onset of Fe-rich regions due to intermixing and would lead to changes in the relative intensity of the different features in the Fourier transform, as it is present for example in ion irradiated Fe$_{60}$Al$_{40}$ thin films \cite{Torre2018}. This is in contrast to XRD measurements \cite{Cervera2017}, where intermixing was suggested due to a decrease of the long-range order parameter $S$ with rising ion irradiation fluence. In fact, the spectra of the disordered sample can be simulated by applying a static mean square relative displacement (MSRD) $\sigma^2_{stat}$ of $9\cdot 10^{-4}\mathrm{\,\mathring{A}^{-2}}$ to the spectra of the ordered (as-grown) sample, while changes of the dynamic contributions can be neglected due to the low measurement temperature (T = 5\,K). This shows that after light ion irradiation of B2-FeRh, an increase of the MSRD is observed. This can be due to the increased structural disorder and lattice distortions, for example, by trapped Ne ions inside the lattice \cite{Bennett2017} or a decreased grain size, increasing the static disorder. DB-VEPAS reveals such an increase of the defect concentration leading to an increase of open volume. Further investigation, for example by positron annihilation lifetime spectroscopy may reveal the defect type.    
In temperature-dependent magnetisation measurements, a ferromagnetic ordering at low temperatures is present, while this is accompanied by a broadening of the hysteresis and a decrease of $T_{tr}$. From the field and temperature-dependent measurements, it is seen that a coexistence of two diffe\-rent magnetically ordered states occurs. The first magnetic phase is a soft ferromagnetic phase and can be attributed to the disorder-induced phase, while the second phase can be described to the initial hard magnetic B2 phase with an AFM ordering, where an increased field breaks the anti-parallel spin alignment and induces the formation of FM ordering. A similar increase of the magnetisation at low temperatures has been observed in different investigations, where, for higher irradiation doses \cite{Heidarian2015,Cervera2017,Fujita2010} a suppression of the AFM-FM transition and even a suppression of the ferromagnetic ordering was observed,
while creating the paramagnetic A1 phase (fcc) . The non-systematic increase of the high field magnetisation as a function of the ion fluence is due to the fact, that the two samples irradiated with 0.05\,Ne$^+$/nm$^{2}$ and 0.075\,Ne$^+$/nm$^{2}$ (afterwards irradiated with 0.05\,Ne$^+$/nm$^{2}$) have a different amount of a secondary A1-FeRh phase (as discussed before concerning the microscopic magnetic structure). The secondary phase has not been considered in the magnetisation normalisation due to the unknown composition of the secondary phase\cite{Swartzendruber1984}.
Defining the remanent macroscopic magnetisation $M_r$ as an order parameter of the phase transition in the temperature-driven or ion-irradiation induced AFM-FM transition and comparing the dependence of $B_{hf}$ as a function of the remanent magnetisation, a similarity between the two distinct phase transitions is present (shown in Figure \ref{fig:Contour-Plots}c and \ref{fig:Contour-Plots}d). For the temperature-driven phase transition for small $M_r$ a decrease of $B_{hf}$ is present, while a secondary magnetic contribution occurs at 10 emu/cm$^3$ from the ferromagnetic phase with the coexistence of both $B_{hf}$ contributions up to 100 emu/cm$^3$. From the temperature dependence for elevated temperatures using a Brillouin function one obtains a hyperfine splitting $B_{hf}=27.4\,\mathrm{T}$ at 300\,K, which corresponds to the hyperfine field splitting of FM Fe$_{51}$Rh$_{49}$ at 300\,K\cite{Shirane1963}. In the ion irradiated samples, an additional magnetic contribution occurs at 27.4\,T, whose spectral contribution increases with rising ion fluence and $M_r$. The occurrence of this ferromagnetic phase can be explained, for example, by a change of the chemical composition and the formation of Fe-rich region (Fe nucleus on a Rh-site). Such a change is not pervasively present in the M\"ossbauer spectra (maximum 0.6\,Fe-at.\% ), consistent with the EXAFS measurement. 
Therefore, the formation of the ferromagnetic ordering needs to be explained by a different process. 
In the thermally induced phase transition a discontinous increase of the short-range disorder (MSRD \cite{Wakisaka2015} or Lamb-M\"ossbauer factor\cite{Wolloch2016}) occurs, while in the ion irradiated sample an increase of the static disorder can be seen, indicated by the increased MSRD $\sigma^2$ (EXAFS).
This is due to induced grain boundaries or lattice distortions owing to ion irradiation. This increase of the defect concentration was confirmed in the positron annihilation spectroscopy (decreasing $L+$). Combining the observed changes of the hyperfine field splitting towards an identical splitting in the FM phase at 300\,K and an increase of the static disorder in EXAFS, while changes of the chemical composition are absent -- the effect can be explained in such a way that a structural defect breaks the local symmetry and modifies the electronic structure, therefore, effectively lowering the transition temperature of the surrounding. This concept of a defect-driven domain nucleation growth in B2 FeRh was previously suggested by Keavney et. al\cite{Keavney2018}. 

In summary, we found an increase of the static structural disorder in FeRh thin films irradiated with low ion fluences, while no substantial intermixing of Fe and Rh atoms can be observed on the local scale by microscopic probes such as CEMS or EXAFS. Besides, PAS-measurements show an increase of defect concentration, while the size of the defects does not change. \isotope[57]Fe-CEMS measurements reveal a second magnetic phase with an increased relative spectral area with rising irradiation fluence. The magnetic properties are identical to the ferromagnetic B2-FeRh system. By comparing the thermal-driven and the structural disorder-driven phase transition a jump-like behaviour of the remanent magnetization dependence of the hyperfine field can be observed. Based on these findings, it may be possible to induce structural defects by varying the microstructure, for example, due to variations in the volume fraction of the A1-phase\cite{Chirkova2017} and, therefore, utilizing the defect-driven nucleation of ferromagnetic domains, as it was suggested in the work of Keavney et al. \cite{Keavney2018}. These findings show a strong correlation between the lattice and the magnetic structure in FeRh thin films and the possibility to tailor the phase transition by defect engineering.

\section{Experimental}

FeRh thin films were grown by molecular-beam epitaxy (MBE) by co-deposition of \isotope[57]Fe-metal (95\% enriched in the isotope \isotope[57]Fe) and Rh in ultrahigh vacuum ($p_{growth}=4\cdot 10^{-9}$\,mbar) on a MgO(001) substrate. Before the deposition of the film, the MgO(001) substrate was cleaned using isopropanol and heated at 300\,$^{\circ}$C for 60\,min in a pressure of $1\cdot 10^{-9}$\,mbar to remove contaminants from the surface structure. During the deposition, the temperature of the substrate was 300\,$^{\circ}$C, while the deposition rates of \isotope[57]Fe and Rh were measured and controlled by independent quartz-crystal oscillators. After deposition the film was in situ annealed at 700\,$^{\circ}$C for 90\,min to ensure the formation of the B2 structure. The thickness of the sample was confirmed to be 42.8\,nm by X-ray reflectivity. Furthermore, the application of a capping layer to prevent oxidation has been avoided based on the observations made in different works\cite{Baldasseroni2012,Baldasseroni2014,Bennett2017,Odkhuu2018}, where the capping layer influenced the magnetic properties of the B2-ordered system in such a way that a ferromagnetic phase was observed at the interface between the film and capping layer. Based on the performed CEMS and PAS measurements, we can neglect the formation of an oxide layer, as will be discussed later. The as-grown sample was cut into three pieces, while the first piece was used to perform magnetometry and temperature-dependent \isotope[57]Fe-CEMS measurements, the second and third sample pieces were used for an initial irradiation with 0.05 Ne$^+$/nm$^2$ and 0.075 Ne$^+$/nm$^2$ respectively. Both irradiated samples were then further irradiated with 0.05 Ne$^+$/nm$^2$ to achieve a total irradiation of 0.1 and 0.125 Ne$^+$/nm$^2$ respectively. In addition, a sample with an irradiation dose of 0.4 Ne$^+$/nm$^2$ was used for EXAFS measurements. For the different irradiation steps an ion energy of 25\,keV has been chosen based on SRIM \cite{SRIM2010} simulations to fully penetrate the film volume while minimizing defects in the substrate.

\noindent
Magnetic characterization was performed using the vibrating sample magnetometer (VSM) option of a Quantum Design PPMS DynaCool, providing magnetic fields up to $\pm$ 9\,T applied parallel to the film surface in a temperature range between 4.3\,K and 400\,K. Temperature-dependent magnetization measurements were performed in an external field of 10\,mT using the ZFC-FC protocol. To achieve a temperature of 450\,K a ceramic sample holder and the VSM oven option (temperature range between 300 and 1000\,K) was used, resulting in an offset at 300\,K due to different sensitivity ranges.

$^{57}$M\"ossbauer spectroscopy at perpendi\-cular incidence of the $\gamma$-rays onto the film surface was performed by detection of conversion electrons. For the detection of the electrons, the sample was installed in a proportional gas counter, i.e. housing with a continuous He gas flow mixed with 4\% CH$_4$ to avoid ionization processes. For the measurement, a constant acceleration M\"ossbauer driving unit was used with a $^{57}$Co source embedded in an Rh matrix, while the velocity of the spectrometer was calibrated with a $\alpha$-Fe foil reference sample at room temperature. The experimental spectra were evaluated by a least-squares fitting routine using the $Pi$ program package \cite{PiLink}. 

Element-specific EXAFS measurements have been per\-formed  at the XAS beamline BM23 at the ESRF \cite{Mathon:vv5115}. Spectra have been recorded at the Fe K absorption edge in the total fluorescence yield detection mode at 45 degrees incident geometry using an energy dispersive detector to detect signal originating from the Fe K$_{\alpha}$-radiation. 

Doppler broadening variable energy positron annihilation spectroscopy (DB-VEPAS) measurements have been conducted using the apparatus for in situ defect analysis \cite{Liedke2015} of the slow positron beamline \cite{Anwand2012} located at the Helmholtz-Zentrum Dresden-Rossendorf. Positrons were extracted from a radioactive \isotope[22]Na source, moderated down to several eV and magnetically guided to the accelerator unit, where they were subsequently accelerated in discrete voltage values in the range of E$_p$ =0.04-35\,keV. This positron energy allows depth profiling of the films from the surface down to about 2\,$\mathrm{\mu m}$ for FeRh, while a mean positron implantation depth can be approximated by a simple material density-dependent formula: $z_{mean}=3.69\cdot E_p^{1.62}$, where $z_mean$ is expressed in the units of nm. The broadening of the 511\,keV annihilation line has been measured with a high purity Ge detector, having energy resolution of 1.09 $\pm$ 0.01\,keV at 511\,keV. Implanted into a solid, positrons lose their kinetic energy due to thermalisation and after short diffusion time, annihilate in delocalized lattice sites or localise in vacancy like defects and interfaces usually emitting two anti-collinear 511\,keV gamma photons once they interact with electrons. Since at the annihilation site thermalised positrons have minimal momentum compared to the electrons a broadening of the 511\,keV annihilation line is observed mostly due to the momentum of the electrons. The broadening of the positron annihilation line is characterised by two distinct parameters S, and W defined as a fraction of the annihilation line in the middle ($511\pm0.93\,\mathrm{keV}$) and outer regions ($508.56\pm0.35\,\mathrm{keV}$ and $513.44\pm0.35\,\mathrm{keV}$), respectively.

\begin{acknowledgement}

The authors thank the Ion Beam Center at Helmholtz-Zentrum Dresden-Rossendorf for providing the necessary facilities for the ion irradiation of the samples. The Impulse- and Networking Fund of the Helmholtz-Association (FKZ VH-VI-442 Memriox), and the Helmholtz Energy Materials Characterization Platform (03ET7015) are acknowledged. Furthermore, the authors thank the European Synchrotron Radiation Facility for provision of synchrotron radiation facilities and allocation of synchrotron radiation within the project HC-3916 at the beamline BM23. We want to thank Ulrich von H\"{o}rsten for his outstanding technical assistance and Markus Gruner (both Duisburg-Essen) for fruitful discussions. This work was supported by the Deutsche Forschungsgemeinschaft (DFG) within the projects WE2623/14-1 and BA5656/1-1.
\end{acknowledgement}

\providecommand{\latin}[1]{#1}
\makeatletter
\providecommand{\doi}
{\begingroup\let\do\@makeother\dospecials
	\catcode`\{=1 \catcode`\}=2 \doi@aux}
\providecommand{\doi@aux}[1]{\endgroup\texttt{#1}}
\makeatother
\providecommand*\mcitethebibliography{\thebibliography}
\csname @ifundefined\endcsname{endmcitethebibliography}
{\let\endmcitethebibliography\endthebibliography}{}

\end{document}